# Ultra-wideband Waveguide-coupled Photodiodes Heterogeneously Integrated on a Thin-film Lithium Niobate Platform

Chao Wei[1], Youren Yu[1], Ziyun Wang[1], Lin Jiang[1], Zhongming Zeng[2], Jia Ye[1], Xihua Zou[1], Wei Pan[1], Xiaojun Xie[1,*] and Lianshan Yan[1,*]

[1]Key Laboratory of Photonic-Electric Integration and Communication-Sensing Convergence, School of Information Science and Technology, Southwest Jiaotong University, Chengdu, 611756, China.

[2] School of Nano Technology and Nano Bionics, University of Science and Technology of China, Hefei, 230026, China.

*xxie@swjtu.edu.cn

*lsyan@swjtu.edu.cn

## Abstract

With the advantages of large electro-optical coefficient, wide transparency window, and strong optical confinement, thin-film lithium niobate (TFLN) technique has enabled the development of various high-performance optoelectronics devices, ranging from the ultra-wideband electro-optic modulators to the high-efficient quantum sources. However, the TFLN platform does not natively promise lasers and photodiodes. This study presents an InP/InGaAs modified uni-traveling carrier (MUTC) photodiodes heterogeneously integrated on the TFLN platform with a record-high 3-dB bandwidth of 110 GHz and a responsivity of 0.4 A/W at a 1550-nm wavelength. It is implemented in a wafer-level TFLN-InP heterogeneous integration platform and is suitable for the large-scale, multi-function, and high-performance TFLN photonic integrated circuits.

## Introduction

Integrated photonics is a discipline of integrating devices such as waveguides, lasers, photodetectors on a single chip. Due to fast and efficient transmission and processing of information, integrated photonics holds a significant potential for applications in communication, computing, and sensing. It offers several advantages such as low cost, low power consumption, and scalability[1]. While materials like silicon[2,3], silicon nitride ($Si_3N_4$) [4,5], and indium phosphide (InP)[6,7] have been widely used as the photonics integration platform, lithium niobate (LN) has still gained considerable attention due to its strong electrooptic coefficient, large nonlinearity, and wide transparency window[8]. However, as a photonics integration platform, traditional bulk LN material faces several challenges such as weak mode confinement, large footprint, and reduced nonlinear efficiency[9]. To overcome these challenges, thin films lithium niobate (TFLN) technique has emerged as a solution. Thanks to the large refractive index difference between LN device layer and $SiO_2$ BOX layer, TFLN technique enables tight mode confinement and high nonlinear efficiency, leading to its wide adoption in optical communications, microwave photonics, THz communications, and quantum photonics[10,11].

Followed by the commercial availability of high-quality TFLN wafers and the breakthrough of its fabrication technique, TFLN technique is implemented to develop various optoelectronics components, such as compact and high-performance modulators[12,13], broadband optical frequency combs[14,15], polarization management devices[16], and efficient wavelength converters[17,18]. Most of these TFLN devices use external lasers and photodetectors since the LN material does not natively provide a light source and photodetection. This is one challenge faced by the TFLN technique as a potential universal photonics integrated circuits (PICs) platform[11]. The flip-chip bonding technique was recently applied to integrate the InP distributed feedback (DFB) laser die with a pre-fabricated TFLN modulator chip[19]. Low-loss and high-power

TFLN-InP transmitters were achieved by optimizing the overlap between the respective platform modes. Besides, as an important optoelectronic conversion device, high-performance integrated photodiodes are essential for the TFLN integration platform. Amorphous silicon photodetectors integrated on TFLN platform was achieved within the visible wavelength range[20]. It features a bandwidth of 10 MHz and responsivity of 22 mA/W to 37mA/W over the wavelengths spanning from 635 nm to 850 nm. The photodiodes working in the wavelength range of the optical communication is still needed to be explored.

Heterogeneous integration is an effective technique to integrate active components on the chip for indirect bandgap materials. Several heterogeneously integrated photodiodes have been successfully developed on silicon-based platforms[21-24]. Recently, heterogeneous integration technique was adopted to achieve broadband photodetection on the TFLN platform[25]. A high-performance photodiode with 80 GHz 3-dB bandwidth and 0.6 A/W responsivity at 1550 nm was demonstrated by using die-bonding technique with SU8 as the adhesion layer. While the broadband photodetection on the TFLN platform had been enabled, an ultra-wideband photodiode with 110 GHz 3-dB bandwidth on the TFLN platform is still highly desired to advance the overall performance of the TFLN PICs, considering the state-of-the-art TFLN modulator exhibits a 3-dB bandwidth of 110 GHz[26]. Moreover, TFLN technique finds its absence of a wafer-level integration of active components, which is the major challenge to achieve a massive-scale, multi-function and high-performance TFLN PICs. Therefore, it is of great significance to achieve wafer-level integration of ultra-wideband photodiodes on a thin-film lithium niobate platform.

This study addresses the challenge by wafer-level heterogeneously integrating InP/InGaAs modified uni-traveling carrier (MUTC) photodiode wafer onto the TFLN wafer with pre-defined waveguides and passive components. The MUTC Epi stack featured with p layers down is implemented to simultaneously boost the bandwidth and responsivity. The fabricated waveguide-coupled photodiodes based on the wafer-level TFLN-InP heterogeneous integration technique have a dark current of approximately 1 nA and a responsivity of 0.4 A/W at a wavelength of 1550 nm. The measured 3-dB bandwidths on 50 Ω load reach up to 110 GHz, which is comparable to the bandwidth of the state-of-the-art TFLN modulator. The devices are successfully applied in a four-level pulse amplitude modulation (PAM4) data-receiving system, which demonstrates the potential of the photodiodes on the TFLN platform for the next-generation high-speed transmission systems.

## Results

### Structure and design

Table 1 Epi structure of the MUTC PDs

| Layer | Material | Thickness | Doping |
|---|---|---|---|
| Substrate | InP | 350 μm | Semi-insulating |
| Buffer layer | InP | 500 nm | N+ doping, $5 \times 10^{18}$ cm$^{-3}$ |
| N-contact layer | In$_{0.53}$Ga$_{0.47}$As | 50 nm | N++ doping, $1 \times 10^{19}$ cm$^{-3}$ |
| Cladding layer | InP | 60 nm | N+ doping, $5 \times 10^{18}$ cm$^{-3}$ |
| Sacrifice layer | InP | 40 nm | P+ doping, $1 \times 10^{18}$ cm$^{-3}$ |
| Drift layer | InP | 120 nm | N- doping, $3 \times 10^{16}$ cm$^{-3}$ |
| Cliff layer | InP | 20 nm | N doping, $3 \times 10^{17}$ cm$^{-3}$ |
| Smooth layer | InGaAsP | 10 nm | N- doping, $1 \times 10^{16}$ cm$^{-3}$ |
| Depleted absorption layer | In$_{0.53}$Ga$_{0.47}$As | 20 nm | N- doping, $1 \times 10^{16}$ cm$^{-3}$ |
| graded doped absorption layer | In$_{0.53}$Ga$_{0.47}$As | 100 nm | P doping, $5 \times 10^{17}$ cm$^{-3}$<br>↓<br>P+ doping, $5 \times 10^{18}$ cm$^{-3}$ |
| P-contact layer | InGaAsP | 280 nm | P+ doping, $8 \times 10^{18}$ cm$^{-3}$ |

Table1 shows the Epi layer structure of the MUTC PDs, which was grown by metal-organic chemical vapor deposition (MOCVD) on a semi-insulating InP substrate. The n-doped layers were initially grown on the InP substrate to establish a p-down structure after wafer bonding. In this structure, the absorption region has a total thickness of 120 nm, consisting of graded doped absorption layers of 100 nm ($5 \times 10^{17}$ cm$^{-3}$ → $5 \times 10^{18}$ cm$^{-3}$) and a depleted absorption layer of 20 nm ($1 \times 10^{16}$ cm$^{-3}$). A self-induced electric field in the graded doped absorption layer enhances the transport of the photo-generated electrons[27]. A 20-nm-thick n-doped cliff layer with $3\times10^{17}$ cm$^{-3}$ doping level is incorporated to mitigate the space charge effect[28]. A 120-nm-thick drift layer is chosen to balance the junction capacitance and carrier transit time. To enable electron velocity overshoot, the electric field in the drift region is regulated by inserting a p-doped sacrifice layer[29]. Heavily doped InGaAs and InGaAsP are used as n and p contact layers, respectively. In addition, InGaAsP layer acts as an optical coupling layer to achieve an efficient optical coupling between the active layers and lithium niobate waveguide. After optimizing the thickness of the InGaAsP layer, a thin absorption layer can perform an effective light absorption due to the strong optical coupling. Compared to the n-down structure, our structure can achieve a high responsivity since the absorption layers are close to the LN waveguide. The LN waveguide has a total thickness of 600 nm and a slab thickness of 300 nm. The width of the LN waveguide is designed as 1μm for a low-loss light transmission.

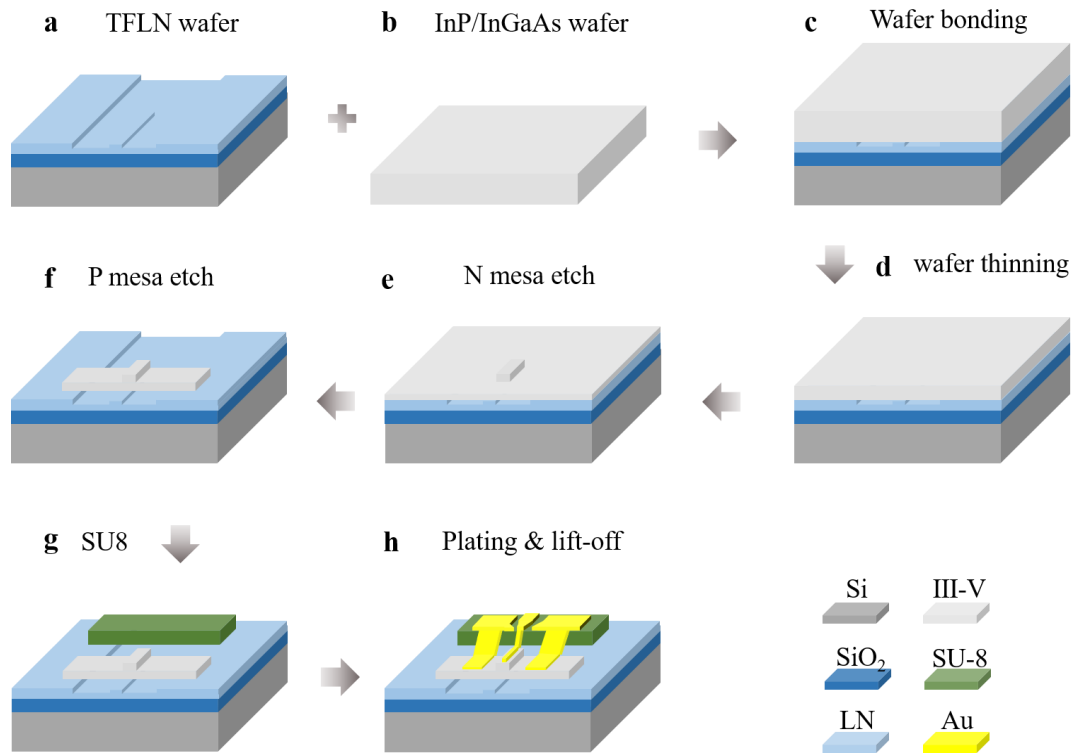

**Fig. 1. The designed process flow used to fabricate the waveguide-coupled photodiodes on the TFLN platform. a** TFLN wafer with pre-defined waveguide and passive components, **b** bare InP/InGaAs wafer, **c** InP/InGaAs wafer and TFLN wafer bonding, **d** InP/InGaAs wafer

substrate removal, **e** N mesa dry etch, **f** P mesa dry etch, **g** SU8 base for CPW pad, and **h** metal electroplating and lift-off.

Figure 1 illustrates the designed fabrication process of the heterogeneously integrated photodiodes on the TFLN wafer, which includes wafer bonding, III-V material etching, and electrode fabrication. Initially, the LN waveguide was obtained by dry etching. The width of the fabricated LN waveguide is approximate 1um. The slope of the waveguide sidewall is approximate 63 degrees. Microscope and scanning electron microscope (SEM) images of the LN waveguide can be seen in Fig. 2a, b, respectively. Subsequently, an InP/InGaAs wafer was bonded onto the TFLN wafer with pre-defined waveguides and passive components. A combination of dry and wet etching techniques is utilized to etch III-V material, forming the n- and p- mesa while avoiding undercut and waveguide damage. The etched n mesa and p mesa is shown in Fig. 2c, d. Finally, a metal layer was electroplated as an electrode shown in Fig. 2e, f. Further details of the fabrication process can be found in the Materials and Methods section. The heterogeneously integrated MUTC PD with active areas of 2 μm × 6 μm, 2 μm × 8μm, 2 μm × 10 μm, 2 μm × 12 μm and 2 μm × 14 μm were fabricated.

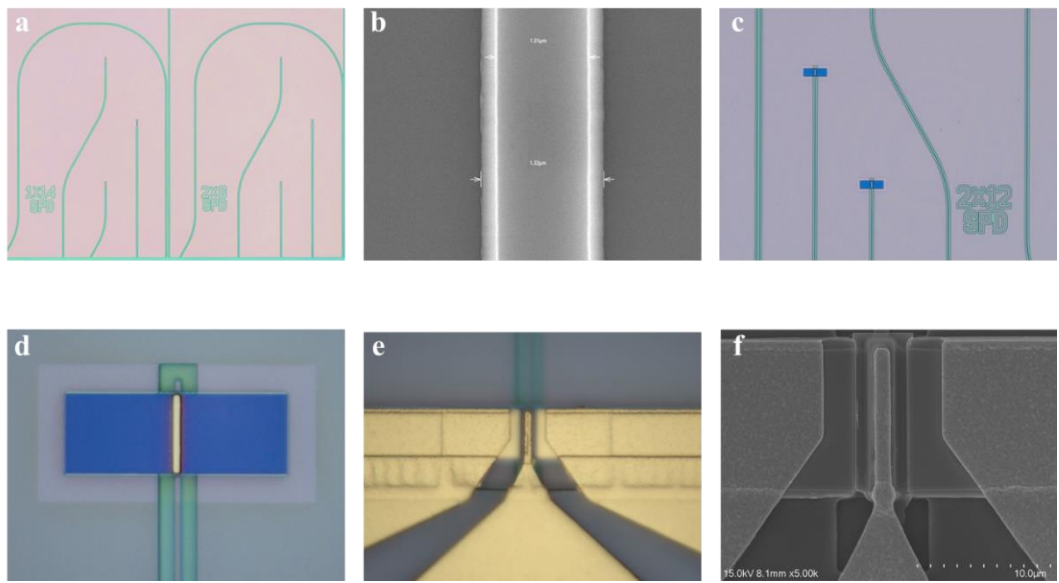

**Fig. 2.** Microscope **a** and SEM **b** images of the LN waveguide, **c, d**, microscope images of the devices after n-mesa and p-mesa etching, microscope **e** and SEM **f** images of the fabricated

photodiodes.

**Measurement results and analyses**

Firstly, we carried out the measurement of the dark current of the fabricated PDs. Figure 3a shows the dark current versus the bias voltage of the devices with active areas of 2 μm × 6 μm ~ 2 μm × 14 μm. The typical dark current of the devices is approximately 1 nA at -4 V bias. Next, the internal responsivity of the waveguide-coupled PDs was characterized. The fiber-to-waveguide coupling loss and waveguide loss were measured to calculate the internal responsivity of the devices. A lensed fiber with a spot size of 2.5 μm was used to couple the light into and out of the loopback waveguides near the target photodiodes. The coupling loss and waveguide propagation loss were approximately 7 dB. The photocurrent was measured by coupling light into the photodiode near the loopback waveguide. At the wavelength of 1550 nm, the internal responsivities of the devices with lengths of 6 ~ 14 μm were 0.4, 0.48, 0.5, 0.54, and 0.55 A/W, considering the measured coupling loss and waveguide propagation loss, as shown in Fig. 3b. The measured responsivities of the devices are consistent with the simulated ones. When the PD length increases, the responsivity saturates due to the intrinsic absorption of the metal stack. The absorption of the metal stack can be mitigated by optimizing the thickness of the cladding layer.

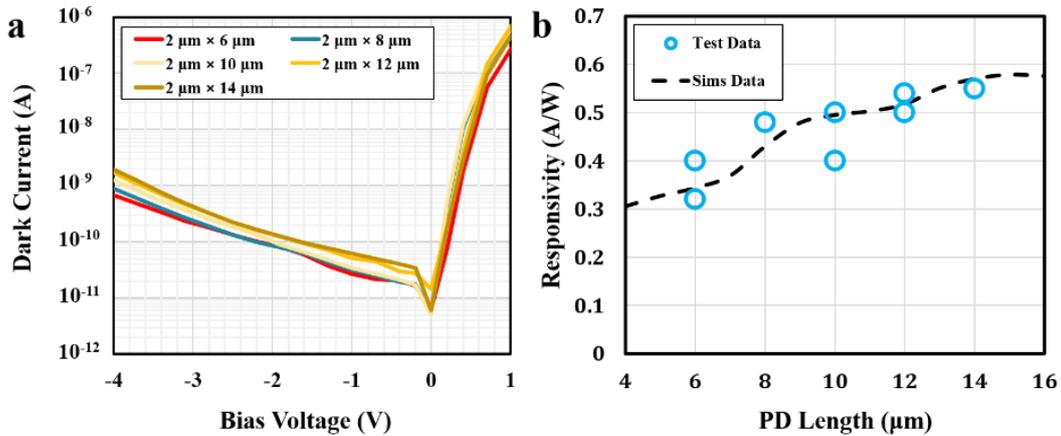

**Fig. 3. a** I-V curves of the devices with active areas of 2 μm × 6 μm ~ 2 μm × 14 μm, **b** measured and simulated responsivities of the devices with lengths of 6 ~ 14 μm.

The S parameter of the devices was measured by a vector network analyzer with a frequency range of DC ~ 67 GHz. By fitting the measured S parameters to the equivalent circuit model as shown in Fig. 4a, the resistance and capacitance of the devices were extracted. The measured and fitted S11 curves of the devices with active areas of 2 μm × 6 μm ~ 2 μm × 14 μm at -4 V bias voltage are presented in Fig. 4b-f. The physical parameters extracted from the measured S11 and the Lumerical 3D model are shown in Fig. 5. The fitted parameters of the devices with different active areas and lengths are consistent with the simulation results. The fabricated photodiodes on the TFLN platform exhibit low junction capacitance but relatively large series resistance which is dominated by the resistance on p-mesa.

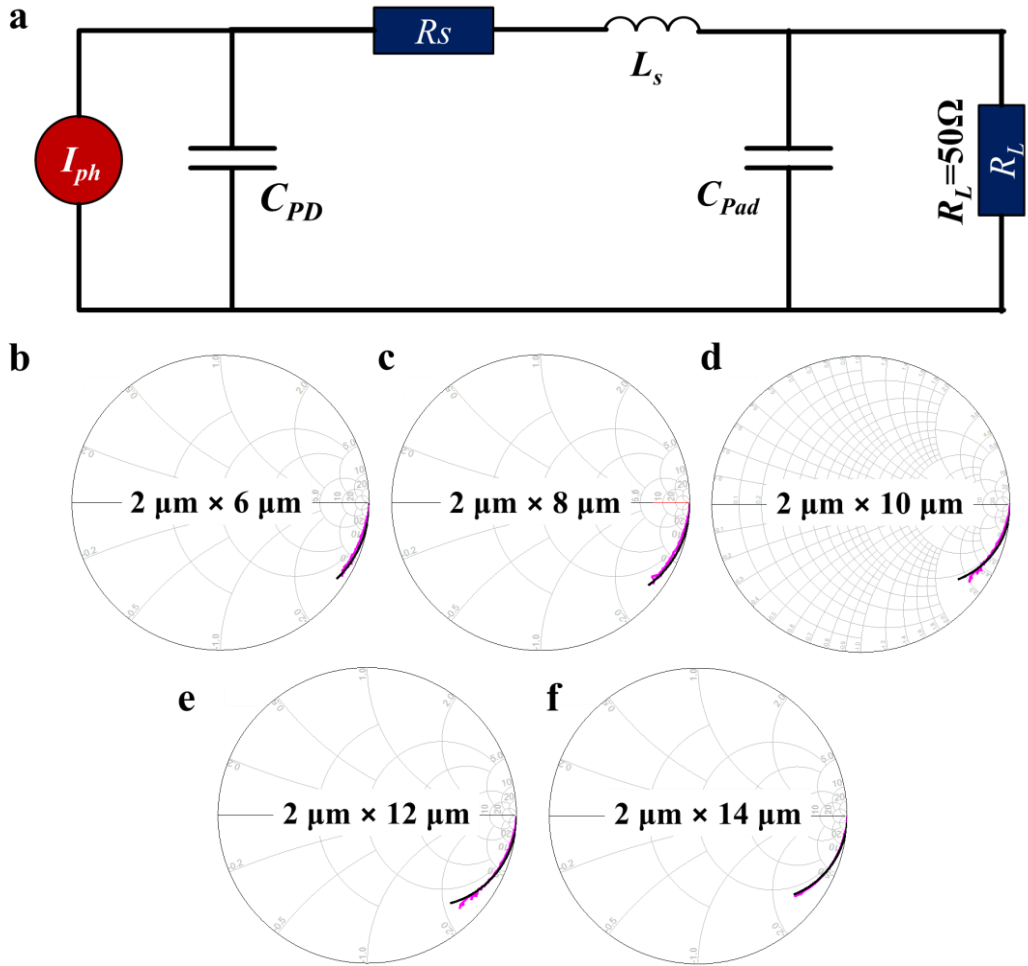

**Fig. 4. a** Equivalent circuit model of the devices for S11 fitting and frequency response simulation. $I_{ph}$, $C_{PD}$, $R_s$, $C_{pad,}$ and $L_s$ represents the frequency-dependent current source, junction

capacitance, series resistance, and the capacitance and inductance of the CPW pattern, respectively. **b-f** Measured (red solid line) and fitted (black solid line) S11 curves (frequency range: 1 ~ 67 GHz) for devices with active areas of 2 μm × 6 μm ~ 2 μm × 14 μm at -4 V bias voltage.

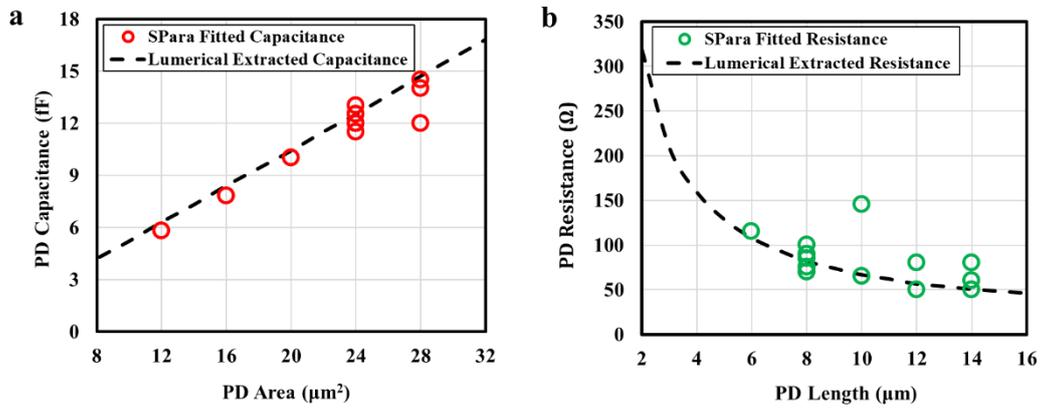

**Fig. 5.** PD capacitance **a** and resistance **b** extracted from the measured S11 parameter and the Lumerical 3D model.

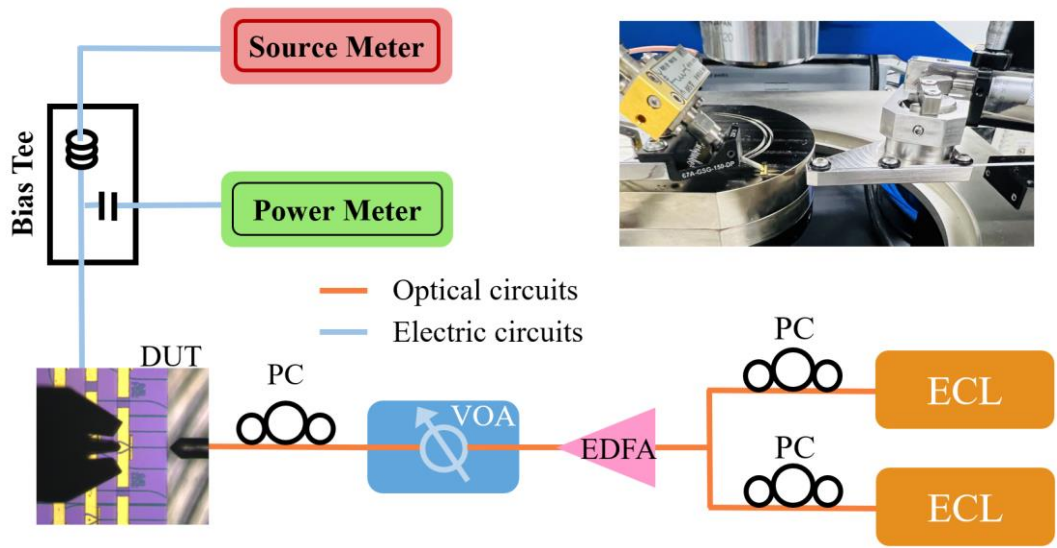

**Fig. 6.** Frequency response measurement setup. ECL: external cavity laser, PC: polarization controller, EDFA: Erbium-doped fiber amplifier, VOA: variable optical attenuator, DUT: device under test. The inset figure shows the experiment setup including bias tee, RF probe, fabricated chip, and lensed fiber.

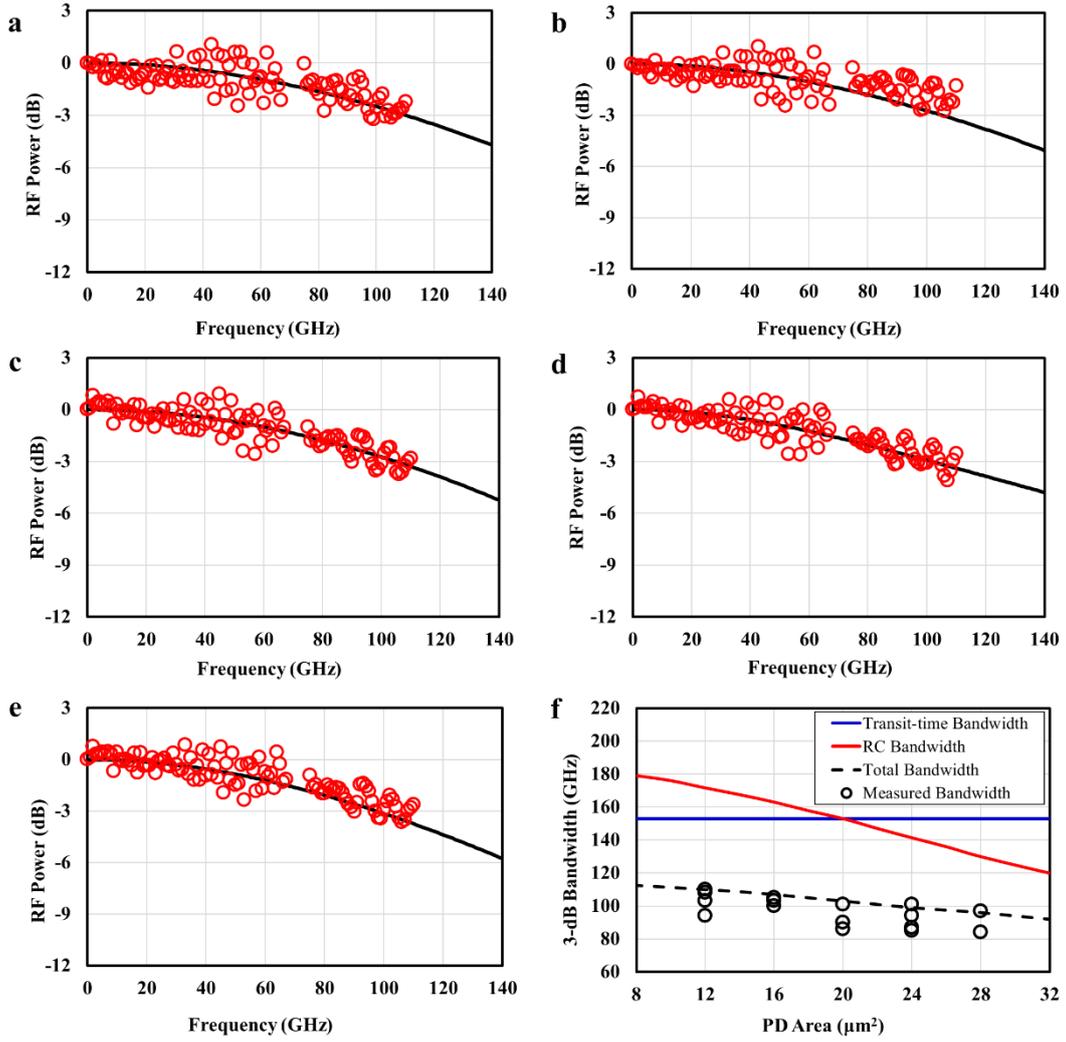

**Fig. 7.** Frequency responses of the devices with active areas of **a** 2 μm × 6 μm, **b** 2 μm × 8 μm, **c** 2 μm × 10 μm, **d** 2 μm × 12 μm, and **e** 2 μm × 14 μm. **f** Transit-time-limited bandwidth (blue solid line), RC-limited bandwidth (red solid line), total bandwidth (black dash line), and measured bandwidth of the devices with various active areas (black circle).

The experimental setup shown in Fig. 6 was used to measure the frequency response of the devices. With RF probes in the frequency ranges DC ~ 67 GHz and 75 ~ 110 GHz, we measured the RF power in the corresponding frequency bands. Figure 7a-e show the power responses of the devices with different active areas. The maximum bandwidths of the devices with active areas of 2 μm × 6 μm, 2 μm × 8 μm, 2 μm × 10 μm, 2 μm× 12 μm, and 2 μm × 14 μm are 110, 105, 100, 101, and 97 GHz, respectively. Figure 7f shows the simulated bandwidths (transit-time bandwidth, RC bandwidth, and total bandwidth) based on physical parameters extracted from the Lumerical 3D model and the measured bandwidths. It can be observed that the measured bandwidths are consistent with the simulated results. The estimated carrier transit time is 2.9 fs and the corresponding transit-time-limited bandwidth is 153 GHz. For the devices with active areas less than 20 $\mu m^2$, the bandwidth is limited by the carrier transit time. When the device area increases and becomes larger than 20 $\mu m^2$, the bandwidth is limited by the RC constant.

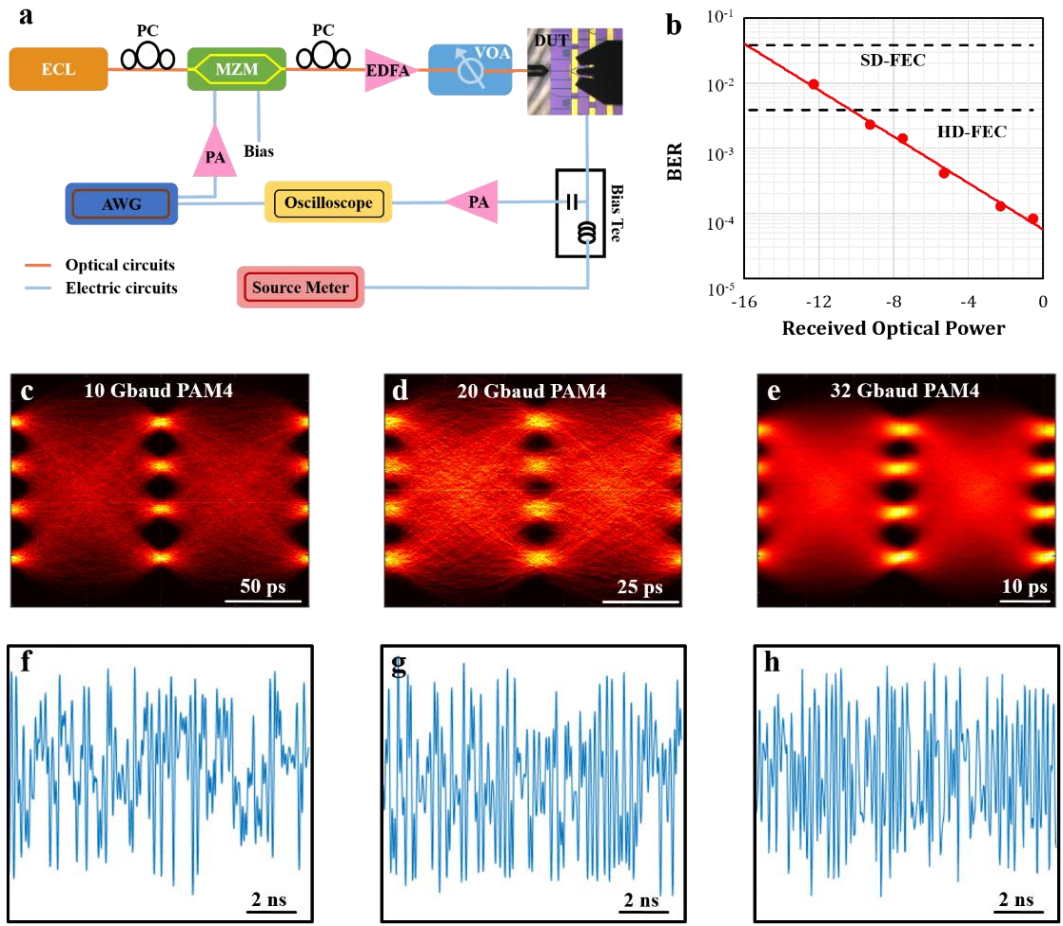

**Fig. 8. a** Intensity modulation and direct detection system setup. ECL: external cavity laser, PC: polarization controller, MZM: Mach-Zehnder modulator, EDFA: Erbium-doped fiber amplifier, DUT: device under test, AWG: arbitrary waveform generator. **b** Measured bit error rates (BERs) versus the received optical power for 32 Gbaud PAM4 signal. Eye diagrams **c-e** and measured waveforms **f-h** of the PAM4 signal with 10, 20, and 32 Gbaud.

To further verify the characteristics of the photodiodes, the devices were applied in the four-level pulse amplitude modulation (PAM4) data-receiving system as shown in Fig. 8a. PAM4 signals with symbol rates of 10, 20, and 32 Gbaud were sent to the devices for demodulation, respectively. The eye diagrams and waveforms of the signals are presented in Fig. 8c-h. The transmission performance of these signals was evaluated by measuring the bit error rates (BERs). 131072 data cycles were used as the transmitting signals and error free transmission at 10, 20, and 32 Gbaud was achieved. In addition, the relationship between the BER and the received optical power of the 32 Gbaud PAM4 signal was studied and shown in Fig. 8b. It can be observed that when the received optical power exceeds -16 dBm, the BER remains below the 20% soft-decision forward error correction (FEC) limit of $2.4 \times 10^{-2}$. Furthermore, when the power exceeds -10 dBm, the BER reaches the levels below the 7% hard-decision FEC limit of $3.8 \times 10^{-3}$. The fabricated photodiode on the TFLN platform shows error-free reception of the PAM4 signals, which demonstrates its potential for the next-generation high-speed transmission systems.

**Discussion**

This paper presents ultra-wideband waveguide-coupled MUTC photodiodes heterogeneously integrated on the thin-film lithium niobate platform. The bandwidths of the devices with active areas of 2 μm × 6 μm, 2 μm × 8 μm, 2 μm × 10 μm, 2 μm × 12 μm, and 2 μm × 14 μm were 110, 105, 100, 101, and 97 GHz, respectively. The bandwidth of the devices is limited by the transit time of the photo-generated carriers. The carrier transit time can be reduced by optimizing the thickness of the absorption layers and incorporating a diffusion block layer. Another factor limiting the device bandwidth is the large series resistance on p mesa. An InGaAs layer can be added as p contact to reduce the contact resistance and thus larger RC-limited bandwidth. By taking these practices, it has the potential to boost the total bandwidth of the devices up to 200 GHz. The measured responsivities of the devices with lengths in the range of 6 ~ 14 μm were 0.4, 0.48, 0.5, 0.54, and 0.55 A/W. The intrinsic absorption of the metal stack sets the upper limit for the responsivity of the devices. An optimized thicker cladding layer with a moderate doping level could mitigate the absorption of the metal stack and achieve a higher responsivity. The devices were applied to a data transmission system, and the obtained results showed their ability to detect 32 Gbaud PAM4 signal with high quality. It is demonstrated that the heterogeneously integrated photodiodes on TFLN platform have the potential to be applied in the next-generation high-speed transmission systems. This work paves the way to achieving massive-scale, multi-function, and high-performance TFLN photonic integrated circuits. Moreover, it holds a great promise for enabling ultra-high-speed optical communications, high-performance integrated microwave photonics, and multi-function integrated quantum photonics.

## Materials and Methods

### Device fabrication

The whole fabrication process of the heterogeneously integrated photodiode starts from patterning LN waveguide and passive components by dry etching. A careful wet clean process is followed to ensure the cleanness of the TFLN wafer. Subsequently, a bare InP wafer was bonded onto the pre-patterned TFLN wafer. After wafer bonding, the InP substrate was mechanically thinned, followed by wet etching to remove the remaining part and expose the heavily doped InGaAs n-contact layer. N-mesa etch was performed to stop precisely at the InGaAsP p-contact layer by combining dry etching and selective wet etching. This step defined the size of the active region. The self-alignment method was applied in the fabrication process to ensure a precise alignment between the active region and the LN waveguide. P-mesa etching was performed using dry etching and selective wet etching to remove the InGaAsP above the LN wafer. In order to ensure the surface cleanness of the exposed LN wafer region, a wet chemical clean process is performed. A 3-μm-thick SU-8 layer was deposited to form the base for the CPW pad. Metal electrodes with a GSG pad were formed by electroplating and lift-off. The wafer was finally diced into small chips and then the chips were laterally polished.

### Device measurement

In the frequency response test system (Fig. 6), a wavelength-tunable optical beat signal with 100% modulation depth was obtained by the heterodyne method, in which the wavelength of one laser was fixed while that of the other one was tuned. The polarization states of the two optical signals were controlled by two polarization controllers (PCs). An erbium-doped fiber amplifier (EDFA) and a variable optical attenuator (VOA) were used to adjust the optical power to the devices. The light was coupled into the waveguide through a lensed fiber with a spot size of 2.5um. The output RF signal of the photodiodes is accessed by the RF probe and sent to the power meter through the bias tee (SHF BT65R). The bias voltage was applied to the GSG pad through the bias tee.

For digital signal demodulation experiment, PAM4 signals with symbol rates of 10, 20, and 32 Gbaud were generated using the arbitrary waveform generator (AWG) with 25 GHz electrical bandwidth and 65 GSa/s sampling rate. The generated PAM4 signals were amplified by a power amplifier (SHF M827 B), and then applied to a 40 GHz Mach-Zehnder modulator (MZM). The output optical signal of the MZM was then amplified by an Erbium-doped fiber amplifier (EDFA) and then detected by the fabricated device. The output RF signal of the device was amplified by a power amplifier (SHF M827 B) and sent to an oscilloscope with 33 GHz electrical bandwidth and 80 GSa/s sampling rate.

**Acknowledgements**

This work was supported by the National Key Research and Development Program (2022YFB2803800); Fundamental Research Funds for the Central Universities (2682022CX025).

**Author Contributions**

X.X. proposed the original concept and designed the devices. C.W. fabricated the device. C.W., Y.Y. and L.J. performed the measurements with the assistance of Z.W. C.W. and X.X. prepared the manuscript. L.J., Z.Z., J.Y., X.Z., W.P., and L.Y. revised the manuscript and made insightful comments. X.X and L.Y. supervised the project.

**Conflict of interest**

The authors declare no competing interests.

# References

1. Zhu, D. et al. Integrated photonics on thin-film lithium niobate. *Adv. Opt. Photonics* **13,** 242-352 (2021).

2. Jalali, B. & Fathpour, S. Silicon photonics. *J. Light. Technol.* **24,** 4600-4615 (2006).

3. Leuthold, J. et al. Nonlinear silicon photonics. *Nat. Photonics* **4**, 535-544 (2010).

4. Roeloffzen, C. G. H. et al. Silicon nitride microwave photonic circuits. *Opt. Express* **21,** 22937-22961 (2013).

5. Moss, D. J. et al. New CMOS-compatible platforms based on silicon nitride and Hydex for nonlinear optics. *Nat. Photonics* **7**, 597-607 (2013).

6. Kish, F. A. et al. Current status of large-scale InP photonic integrated circuits. *IEEE J. Select. Topics Quantum Electron.* **17,** 1470-1489 (2011).

7. Nagarajan, R. et al. InP photonic integrated circuits. *IEEE J. Select. Topics Quantum Electron.* **16,** 1113-1125 (2010).

8. Honardoost, A. et al. Rejuvenating a versatile photonic material: thin-film lithium niobate. *Laser & Photonics Rev.* **14,** 2000088 (2020).

9. Andreas, B. et al. Status and potential of lithium niobate on insulator (LNOI) for photonic integrated circuits. *Laser & Photonics Rev.* **12**, 1700256 (2018).

10. Marpaung, D. et al. Integrated microwave photonics. *Nat. Photonics* **13,** 80–90 (2019).

11. Boes, A. et al. Lithium niobate photonics: Unlocking the electromagnetic spectrum. *Science* **379,** abj4396 (2023).

12. Wang, C. et al. Integrated lithium niobate electro-optic modulators operating at CMOS-compatible voltages. *Nature* **562,** 101–104 (2018).

13. He, M. B. et al. High-performance hybrid silicon and lithium niobate Mach–Zehnder modulators for 100 Gbit s$^{-1}$ and beyond. *Nat. Photonics* **13,** 359–364 (2019).

14. Zhang, M. et al. Broadband electro-optic frequency comb generation in a lithium

niobate microring resonator. *Nature* **68,** 373–377 (2019).

15. He, Y. et al. Self-starting bi-chromatic LiNbO3 soliton microcomb. *Optica* **6,** 1138–1144 (2019).

16. Lin, Z. J. et al. High-performance polarization management devices based on thin-film lithium niobate. *Light Sci. Appl.* **11,** 93 (2022).

17. Chen, J. Y. et al. Ultraefficient frequency conversion in quasi-phase-matched lithium niobate microrings. *Optica* **6**, 1244–1245 (2019).

18. Lu, J. J. et al. Periodically poled thin-film lithium niobate microring resonators with a second-harmonic generation efficiency of 250,000%/W. *Optica* **6,** 1455–1460 (2019).

19. Shams-Ansari, A. et al. Electrically pumped laser transmitter integrated on thin-film lithium niobate. *Optica* **9,** 408-411 (2022).

20. Desiatov, B. & Lončar, M. Silicon photodetector for integrated lithium niobate photonics. *Appl. Phys. Lett.* **115,** 121108 (2019).

21. Xie, X. J. et al. High-power and high-speed heterogeneously integrated waveguide-coupled photodiodes on silicon-on-insulator. *J. Light. Technol.* **34,** 73-78 (2016).

22. Hulme, J. et al. Fully integrated microwave frequency synthesizer on heterogeneous silicon-III/V. *Opt. Express* **25,** 2422-2431 (2017).

23. Wang, Y. et al. High-Power Photodiodes With 65 GHz Bandwidth Heterogeneously Integrated onto Silicon-on-Insulator Nano-Waveguides. *IEEE J. Sel. Top. Quantum Electron.* **24,** 2 (2018).

24. Yu, F. X. et al. High-Power High-Speed MUTC Waveguide Photodiodes Integrated on $Si_3N_4$/Si Platform Using Micro-Transfer Printing. *IEEE J. Sel. Top. Quantum Electron.* **29,** 3 (2022).

25. Guo, X. W. et al. High-performance modified uni-traveling carrier photodiode integrated on a thin-film lithium niobate platform. *Photonics Res.* **10,** 1338 (2022).

26. Xu, M. Y. et al. Dual-polarization thin-film lithium niobate in-phase quadrature modulators for terabit-per-second transmission. *Optica* **9,** 61-62 (2022).

27. Ito, H. et al. High-speed and high-output InP-InGaAs uni-traveling carrier photodiodes. *IEEE J. Sel. Topics Quantum Electron.* **10,** 709–727 (2004).

28. Beling, A. et al. High-power, high-linearity photodiodes. *Optica* **3,** 328-338 (2016).

29. Wei, C. et al. >110 GHz High-Power Photodiode by Flip-Chip Bonding. 2022 IEEE International Topical Meeting on Microwave Photonics (MWP).